\documentclass{elsarticle}
\usepackage{color}

\usepackage{amssymb}
\usepackage{csquotes}

\journal{Nuclear Instruments and Methods A}

\begin{document}

\begin{frontmatter}



\title{Waterproofed Photomultiplier Tube Assemblies for the Daya Bay Reactor Neutrino Experiment}

\author[lble]{Ken Chow}
\author[sc]{John Cummings}
\author[lble]{Emily Edwards}
\author[lblp]{William Edwards}
\author[uiuc]{Ry Ely}
\author[lble]{Matthew Hoff}
\author[uh]{Logan Lebanowski}
\author[sd]{Bo Li}
\author[sd]{Piyi Li}
\author[uh]{Shih-Kai Lin}
\author[uiuc]{Dawei Liu}
\author[ihep]{Jinchang Liu}
\author[ucb,lblp]{Kam-Biu Luk}
\author[sd]{Jiayuan Miao}
\author[tu]{Jim Napolitano}
\author[lblp]{Juan Pedro Ochoa-Ricoux}
\author[uiuc]{Jen-Chieh Peng}
\author[nju]{Ming Qi}
\author[ucb,lblp]{Herbert Steiner}
\author[rpi]{Paul Stoler}
\author[lble]{Mary Stuart}
\author[ihep]{Lingyu Wang}
\author[ihep]{Changgen Yang}
\author[lblp]{Weili Zhong}

\address[ihep]{Key Laboratory of Particle Astrophysics, Institute of High Energy Physics, Beijing 100049, P. R. China}
\address[lblp]{Physics Division, Lawrence Berkeley National Laboratory, Berkeley, CA 94720, U.S.A.}
\address[lble]{Engineering Division, Lawrence Berkeley National Laboratory, Berkeley, CA 94720, U.S.A.}
\address[nju]{Department of Physics , Nanjing University, Nanjing, 210000, P. R. China}
\address[rpi] {Department of Physics, Applied Physics and Astronomy, Rensselaer Polytechnic Institute, Troy, NY 12180, U.S.A.}
\address[sd]{School of Physics, Shandong University, Jinan, 250100, P. R. China} 
\address[sc] {Department of Physics and Astronomy, Siena College, Loudonville, NY 12211, U.S.A.}
\address[tu]{Department of Physics, Temple University, Philadelphia, PA 19122, U.S.A.}
\address[ucb]{Department of Physics, University of California, Berkeley, CA 94720, U.S.A.}
\address[uh]{Department of Physics, University of Houston, Houston, TX 77204-5005, U.S.A.}
\address[uiuc]{Department of Physics , University of Illinois at Urbana-Champaign, Urbana, IL 61801, U.S.A.}

\begin{abstract}
In the Daya Bay Reactor Neutrino Experiment 960 20-cm-diameter waterproof photomultiplier tubes are used to instrument three water pools as Cherenkov detectors for detecting cosmic-ray muons. 
Of these 960 photomultiplier tubes, 341 are recycled from the MACRO experiment. A systematic program was undertaken to refurbish them as waterproof assemblies. 
In the context of passing the water leakage check, a success rate better than 97$\%$ was achieved. 
Details of the design, fabrication, testing, operation, and performance of these waterproofed photomultiplier-tube assemblies are presented.

\end{abstract}

\end{frontmatter}


\section{Introduction}\label{sec:intro}
The Daya Bay Reactor Neutrino Experiment is designed to measure the neutrino-mixing angle $\theta_{13}$ with high precision~\cite{bib:proposal}. 
To reach this goal, the experiment utilizes detectors located in 
one far and two near underground experimental sites 
at different distances from three pairs of reactors. At each site multiple detectors filled with 0.1$\%$ Gd-loaded liquid scintillator are used to detect electron anti-neutrinos from the reactors. Each site is shielded by significant overburden to reduce the flux of cosmic-ray muons, and each anti-neutrino detector (AD) is
submerged in a water pool to suppress ambient 
gamma-ray and neutron backgrounds. Except at the top, each water pool is segmented into two optically isolated zones that are instrumented with 20-cm-diameter photomultiplier tubes (PMTs) as independent water Cherenkov detectors for detecting muons~\cite{bib:muonSys}. \\

In each of the two near sites (EH1 and EH2), the water pool is approximately $16~m\times10~m\times10~m$, and is instrumented with 288 20-cm waterproof PMTs. In the far site (EH3), the water pool is approximately $16~m\times16~m\times10~m$, and is populated with 384 20-cm PMTs. Of the 960 PMTs, 619 are new Hamamatsu R5912 PMTs~\cite{bib:r5912} custom-built as waterproof assemblies for Daya Bay. Since the PMTs in the water pools are used only for tagging incoming muons, we decided to recycle a number of PMTs that had been used in the MACRO (Monopole, Astrophysics and Cosmic Ray Observatory) experiment in Italy~\cite{bib:macro}. We succeeded in waterproofing about 400 MACRO PMTs that passed a pressure test in water, and selected 341 of them 
to instrument the water pools of the Daya Bay experiment.\\ 

In section 2, we discuss the design of the waterproof PMT assemblies, which includes the MACRO PMT, voltage divider, potting shell, and studies of potting materials. In section 3 we describe the potting procedure in detail, including the pressure test used to verify that the potted assemblies were waterproof. Based on the initial pressure test results, the fabrication procedure was improved and the mass production of all $\sim$400 waterproofed PMTs was completed successfully. The waterproofed assemblies that passed the pressure test and a subsequent electrical test were installed into the water pools. Their performance in the water pools during data taking is presented. Section 4 is the summary.\\

\section{Design of the Waterproof PMT Assembly}\label{sec:design}

\subsection{MACRO PMT}\label{sec:macro}
The MACRO PMTs we used are EMI (now Electron Tubes) models 9350KA, and D642KB~\cite{bib:etpmtds}. They have a 20-cm-diameter hemispherical glass window with a blue-green sensitive bialkali photocathode. Each has 11 high-gain SbCs dynodes and 3 BeCu dynodes of 
linear-focused design for good linearity and timing. Since their structure, geometry and light detection features
 are similar to those of the newly purchased Hamamatsu R5912, the 9350KA and D642KB 
 are viable candidates for instrumenting the Cherenkov detectors of the Daya Bay
experiment. 
However, in the MACRO experiment, these PMTs were immersed in mineral oil and their bases were not sealed, whereas, at Daya Bay, they are submerged in ultra-pure water and subject to greater pressures. Thus, it was necessary to encapsulate the electrical components in the bases with epoxy to make them waterproof, and ensure they could withstand sufficient pressure before installing them in the water pools at Daya Bay. 

\subsection{Design of Voltage Divider}\label{sec:base}

In the MACRO experiment, two cables were utilized to operate a PMT, one carrying negative high-voltage (HV) and the other one carrying the signal. However, in the Daya Bay experiment, a single cable carries both positive HV and signal from the base to a decoupler, which isolates the signals from the applied HV and directs them to the front-end electronics. Thus, we designed and fabricated new bases for the MACRO PMTs. Several operational characteristics of the PMTs are sensitive to the base design, such as the working HV at a particular gain, rise and fall times, linearity and collection efficiency. Generally, the specifications of the MACRO PMTs are similar to those of the Hamamatsu R5912s; for example, a maximum gain of 3$\times10^{7}$ at a HV of less than 2 $kV$, and linearity better than $\pm5\%$ at $60~mA$ peak anode-current. \\

\begin{figure}[htbp] 
\centering
\includegraphics[width=18cm,height=5.5cm,angle=90]{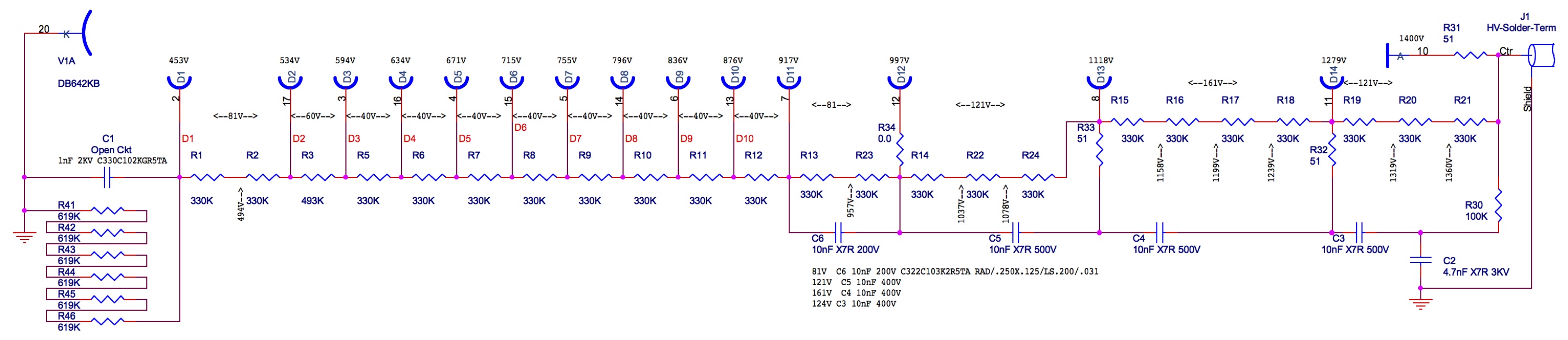} 
\caption{Schematic diagram of the voltage divider for operating EMI 9350KA or D642KB photomultiplier tube with positive high voltage.}
\label{fig:baseDiagram}
\end{figure}

In our design
of the circuit as shown in Figure~\ref{fig:baseDiagram}, the values of the resistors follow the recommendation of the manufacturer of the EMI 9350KA. In particular, the resistor between the photocathode and first dynode provides a potential difference of around $450~V$ at our nominal operating voltage. Our tests have shown that similar performance is achieved for the EMI D642KB PMTs with the same circuit design. 
Based on the evaluation of a few prototypes, we found that the combination of a 50-$\Omega$ terminating resistor (R30) and no damping resistor (R31) at the output of the anode provided an optimal waveform, i.e., it minimized both the recovery time after the overshoot and the ratio of the overshoot amplitude to the signal amplitude. The circuit was carefully laid out
to reduce potential electromagnetic interference between components. 

\subsection{Design of Potting Shell } \label{sec:shell}
\begin{figure}[htbp] 
\centering
\includegraphics[width=45mm, height=60mm]{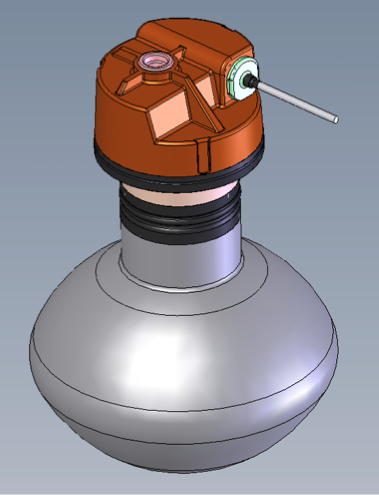} 
\includegraphics[width=60mm, height=60mm]{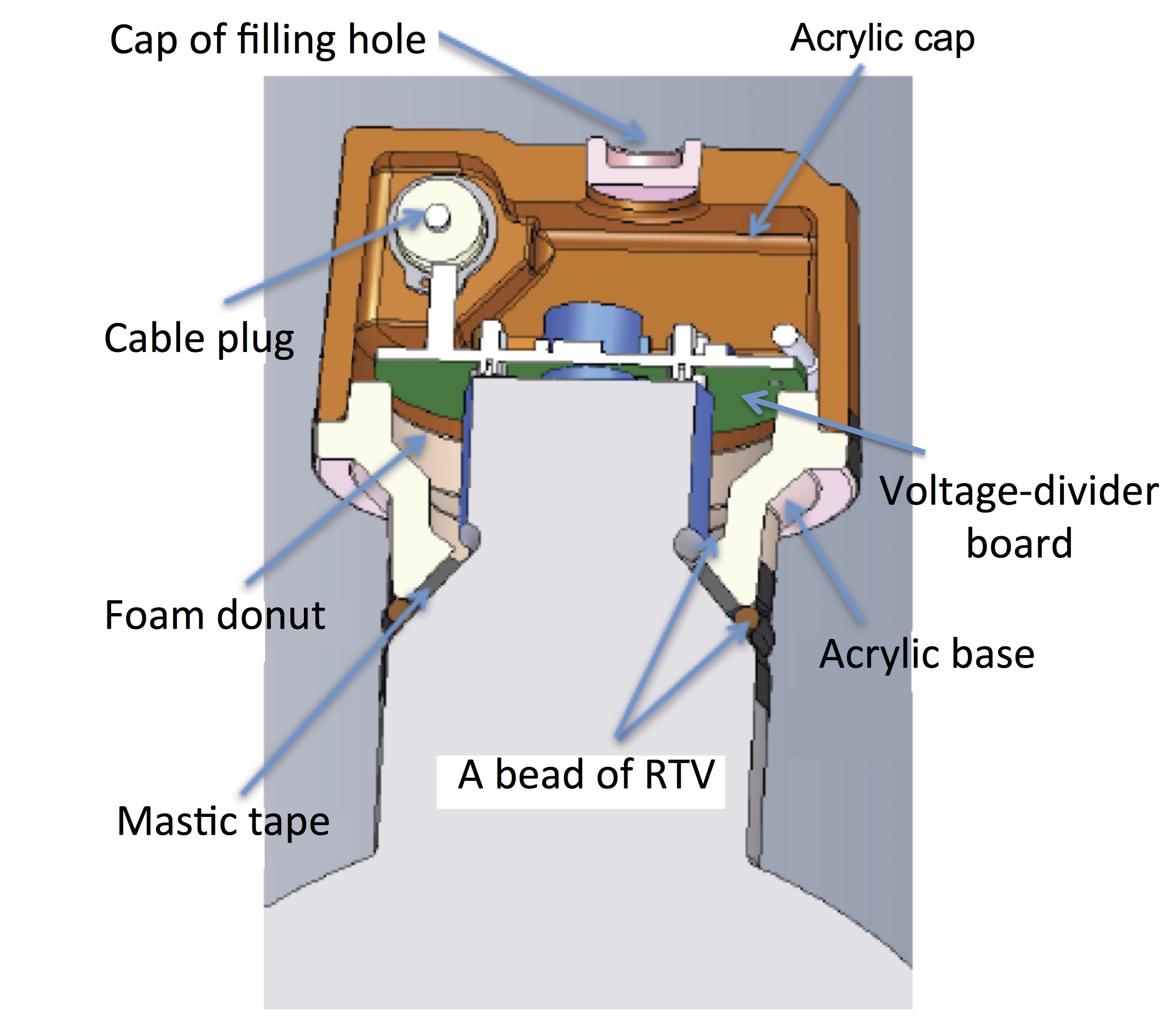} 
\caption{Design of PMT potting shell. Left: schematic diagram of a MACRO PMT with a potting shell. Right: vertical cross-section of the potting shell.}
\label{fig:shellDesign}
\end{figure}
To ensure that the refurbished MACRO PMTs work in water, their bases must be well sealed. To achieve this goal, the potting shell has to have minimal contact between the epoxy and water. Our design shown in the left picture of Figure~\ref{fig:shellDesign} also needs to satisfy the constraint that the potted tube geometry and PMT mounts be compatible with those for the purchased Hamamatsu waterproof assemblies. To achieve this, the profiles of a few MACRO PMTs were scanned and modeled by a coordinate measuring machine (CMM). The tube length, socket diameter and dome diameter of all 400 MACRO PMTs were then measured. The variations of the parameters were within tolerance for mounting the potted PMTs on support frames for installation. The drawing on the right of Figure \ref{fig:shellDesign} is a vertical cross-section of the two-piece potting shell. One piece is the acrylic base that sits on the cone-shaped glass envelope of the PMT and holds the plastic socket with the connecting pins. The voltage-divider board plugs onto the socket. The other piece is the acrylic cap, which joins with the acrylic base. The top of the acrylic cap has a crossing rib design that strengthens the potting shell. The side wall of the acrylic cap has an opening that allows the coaxial cable to exit the base. Another opening in the top of the acrylic cap (labeled as \enquote*{filling hole} in Figure \ref{fig:shellDesign}) is for pouring epoxy into the volume between the two acrylic pieces to seal the PMT base. At the narrow end of the acrylic base, there are four layers of protection from inside to outside: epoxy, sealant, mastic tape and sealant. At the acrylic cap, the coaxial cable also has four layers of protection: epoxy, sealing plug, sealing acrylic cap and mastic tape. The acrylic base and acrylic cap are sealed together with acrylic cement. The filling hole in the top of the acrylic cap is closed by a smaller acrylic cap and sealed with the same acrylic cement. Thus, whichever side of the potting shell is seen by water, there are multiple layers of protection to stop water from entering the potting shell and reaching the electrical connections.

\subsection{Potting Materials}\label{sec:mate}
\subsubsection{Epoxy}
Epoxy is the most important ingredient of the potting materials. It plays the key role in sealing quality and must be effective. It must be compatible with the electrical circuitry of the base, the PMT glass and the cable jacket. The following criteria were considered when choosing the epoxy: \\
\begin{enumerate}
\item suitable for casting, and with low viscosity;
\item curing at room temperature or below; 
\item good bonding with acrylic, polyethylene (PE) and glass;
\item exothermal temperature rise during curing not to exceed 105$^\circ$C (PE limit);
\item suitable for protecting electronics against long term submersion in ultra-pure water at 23$^\circ
$C.
\item acceptable amount of natural radioactivity.
\end{enumerate}

\begin{figure}[htbp] 
\centering
\includegraphics[width=40mm, height=45mm]{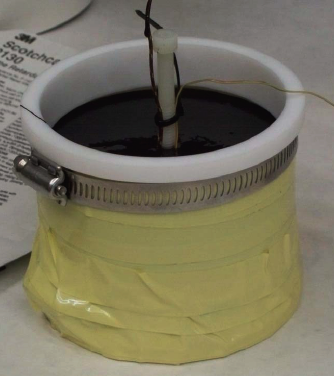} 
\includegraphics[width=40mm, height=45mm]{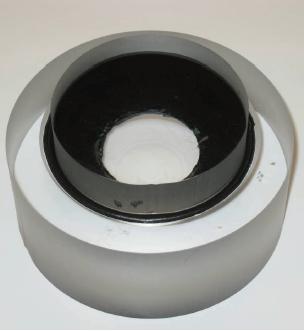} 
\caption{Evaluation of epoxy and mastic tape. The left image is the exothermal test of 3M Scotch Cast SC2130 epoxy. The right image shows the adhesion of Scotch Seal 2229 mastic tape to an acrylic potting base}
\label{fig:epoxyAndMasticTape}
\end{figure}
Based on the above criteria, three candidates were selected for testing: Master Bond EP42-2LV, Master Bond EP30M3LV and 3M Scotch Cast SC2130. 3M Scotch Cast SC2130 was chosen for its low viscosity, lower surface and center curing temperatures, slower hardening and softer \enquote*{feel} (which is believed to indicate greater elasticity), and the absence of temperature spikes in the casting tests. The procedure of the exothermal test consisted of pouring epoxy into a mold and placing a temperature sensor just below the surface at the center (Figure \ref{fig:epoxyAndMasticTape}). The maximum temperature of the 3M Scotch Cast SC2130 within 30 minutes was $61^\circ$C. The amount of natural radioactivity of the 3M Scotch Cast SC2130 was determined to be $38\pm10~ppb$ from the $^{238}$U series, $51\pm15~ppb$ for the $^{232}$Th sequence, and ($0.24\pm0.01)\%$ for natural potassium. It is low enough that this source of radioactive background does not impact the performance of the water Cherenkov detectors. 

\subsubsection{Mastic Tape}
Mastic tape is another important sealing material. It bonds the acrylic base to the cone-shaped part of the glass envelope of the PMT. It is also used to seal the space between the coaxial cable and the opening of the acrylic cap. 
3M Scotch Seal 2229 was selected for its following merits:
\begin{enumerate}
\item Malleable, durable, tacky mastic;
\item Excellent adhesion and sealing to synthetic cable insulations;
\item Very low cold-flow;
\item Used in sealing high-voltage cable splices;
\item Seals against dust, soil and water.
\end{enumerate}
Radioactivity of the Scotch Seal 2229 mastic tape was determined to be $1.78~ppm$ ($^{238}$U series), $4.5~ppm$ ($^{232}$Th sequence) and 0.142\%
(natural K). Since only a small amount of this tape is used in each assembly, the radioactive contaminants are again not a concern.

\subsubsection{Marine sealant and foam donut}
Additional sealing materials were also used to mitigate leakage.
One such material is the 3M Marine Adhesive Sealant Fast Cure 5200, which is normally used in marine underwater applications. The marine sealant was applied inside and outside the acrylic base around the mastic tape, as shown in the right drawing of Figure \ref{fig:shellDesign}, where it is labeled as \enquote{RTV}.\\

\begin{figure}[htbp] 
\centering
\includegraphics[width=45mm, height=60mm]{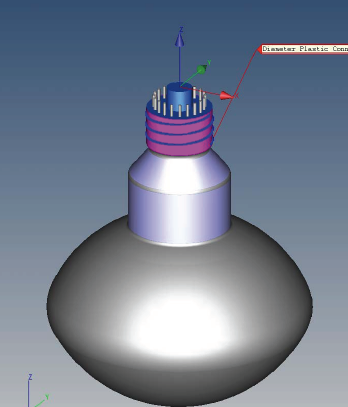} 
\includegraphics[width=65mm, height=60mm]{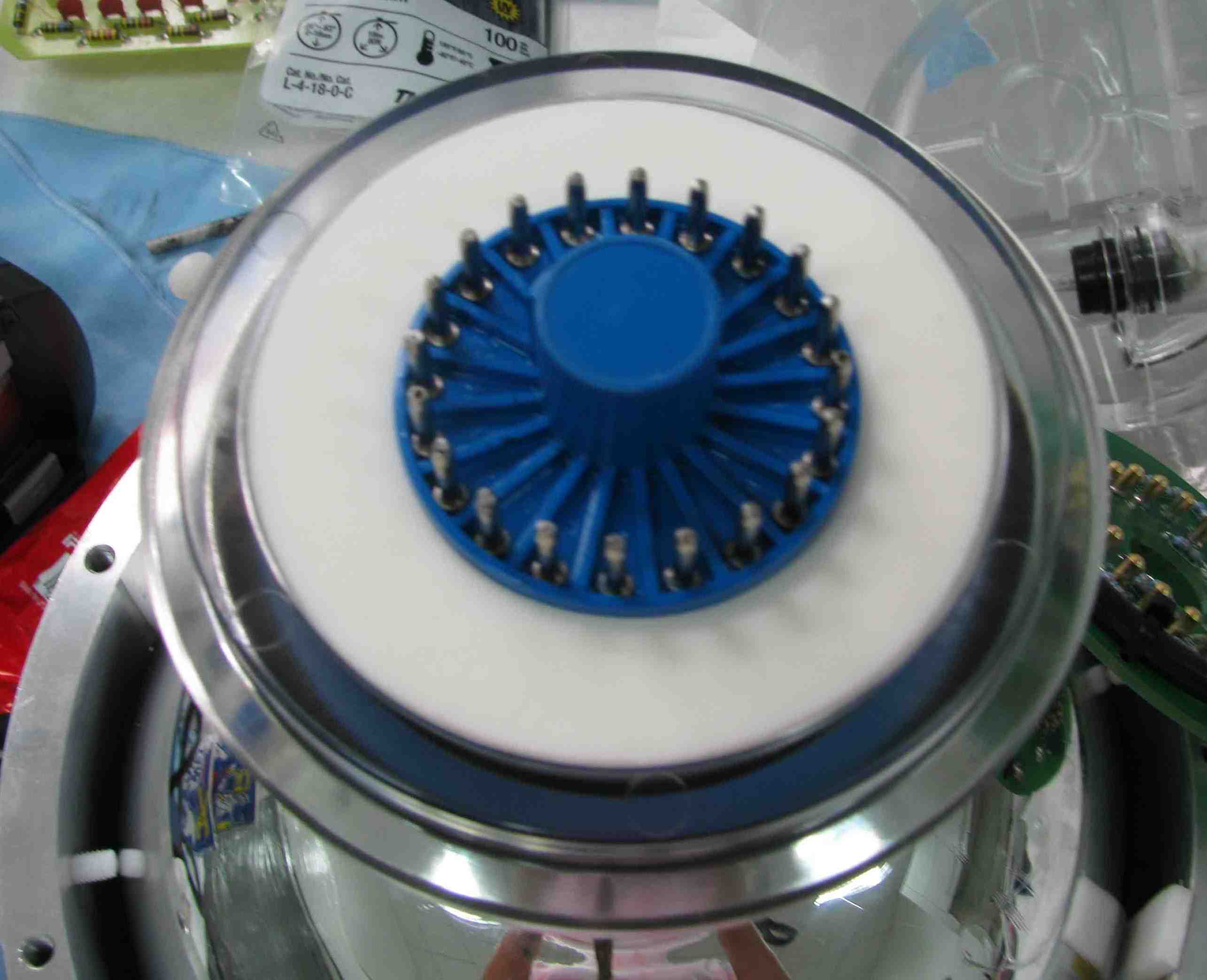} 
\caption{Left: A CMM scan showing the tilt between the axis of the cone-shaped PMT bottom and the axis of the PMT socket with connecting pins. Right: 
A foam donut placed in the clearance between the socket and 
the acrylic base.}
\label{fig:foamDonut}
\end{figure}

The CMM results of 14 MACRO PMTs gave a variation in the cylindricality of the blue socket from $100~\mu m$ to $280~\mu m$. Referring to the 
left drawing of Figure \ref{fig:foamDonut}, the distance between the centerlines of the socket and the cone-shaped bottom of the PMT varied with an average of $1.8~mm$  (with the largest difference being $3.4~mm$), which means there is a slight angle between the two centerlines. Given this, we first applied the marine sealant to the socket connector. Then, because of the angle, we added a PE foam donut in the clearance between the acrylic base and socket (see the right image of Figure \ref{fig:foamDonut}), and sealed around the mastic tape inside the acrylic shell with the marine sealant.

\subsubsection{Cable Sealing}

\begin{figure}[htbp] 
\centering
\includegraphics[width=100mm, height=50mm]{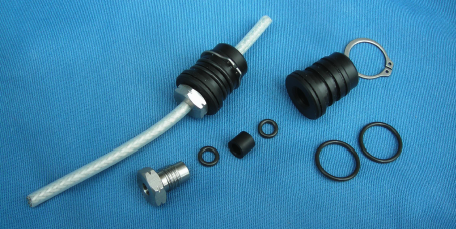} 
\caption{A UW-PSL sealing plug and its components.}
\label{fig:UWplug}
\end{figure}
The cable carrying the signal and HV is a 52-m-long 50-$\Omega$ 
coaxial cable (model number C07947E) made by JUDD Wire Inc., with the outer jacket made of high-density polyethylene \cite{bib:judd}.
As displayed in Figure \ref{fig:shellDesign} there is an opening in the side wall of the acrylic cap for the coaxial cable to penetrate. A steel/plastic sealing plug that is custom-made by the University of Wisconsin-Madison Physical Sciences Lab (UW-PSL) was chosen to create the seal between the cable and the opening in the acrylic cap, as shown in Figure \ref{fig:UWplug}. The diameter of the sealing plug matches the size of the opening of the acrylic shell. Two O-rings seal the plug-cap interface and two smaller O-rings seal the cable inside the sealing plug. To further protect the base from water entering along the coaxial cable, the cable jacket was etched to improve its adhesion to the epoxy. Chemical etching and plasma etching were tested, with the latter approach yielding better results. After etching about $20~cm$ of the jacket, the cable was dipped into the Scotch Cast SC2130 potting compound, and cured overnight before trimming and soldering it onto the voltage-divider board.  

The acrylic shells, epoxy, mastic tape, marine sealant, and cable sealing plug were all compatible with ultra-pure water. 

\section{Mass Production of Waterproof PMT Assemblies}\label{sec:production}
To complete the base sealing of about 400 PMTs, a potting laboratory was set up for mass production.  The laboratory was partitioned into specialized areas for cleaning, assembling, potting, and mechanical and electrical acceptance testing. The potting of all 400 PMTs lasted for about six months and was completed in March 2010.

\subsection{Screening test}\label{pretest}
Before cleaning, each MACRO PMT was tested with a removable base in 
a dark box.  Its noise signals were viewed with an oscilloscope and
were required to be consistent with those due to single photo-electrons. 
PMTs that failed the check were rejected.
In addition, the integrity of the coaxial cable was checked. 

\subsection{Cleaning}\label{cleaning}
Components that passed visual inspection were cleaned with detergent (Alconox) dissolved in 60$^\circ$C water in an ultrasonic cleaner for 10 minutes. The potting components were thoroughly brushed and rinsed in ultra-pure water.  They were then cleaned again with
60$^\circ$C ultra-pure water in another ultrasonic cleaner for 10 minutes. All cleaned acrylic potting components were baked at 60$^\circ$C in an oven for about six hours. Once baking was done, visual inspection of the parts was performed again before assembling them onto a cleaned PMT. The voltage-divider boards were cleaned with ethanol at 40$^\circ$C in an ultrasonic cleaner for 3 minutes. 

The MACRO PMTs were also carefully cleaned. The glass surface was gently wiped with fiberless cloths. The plastic socket usually contained some mineral oil left behind from the MACRO experiment. To remove the mineral oil, the socket was submerged in ethanol for 5 to 30 minutes, depending on how much oil was present. \\  

\begin{figure}[htbp] 
\centering
\includegraphics[width=60mm, height=45mm, angle=270]{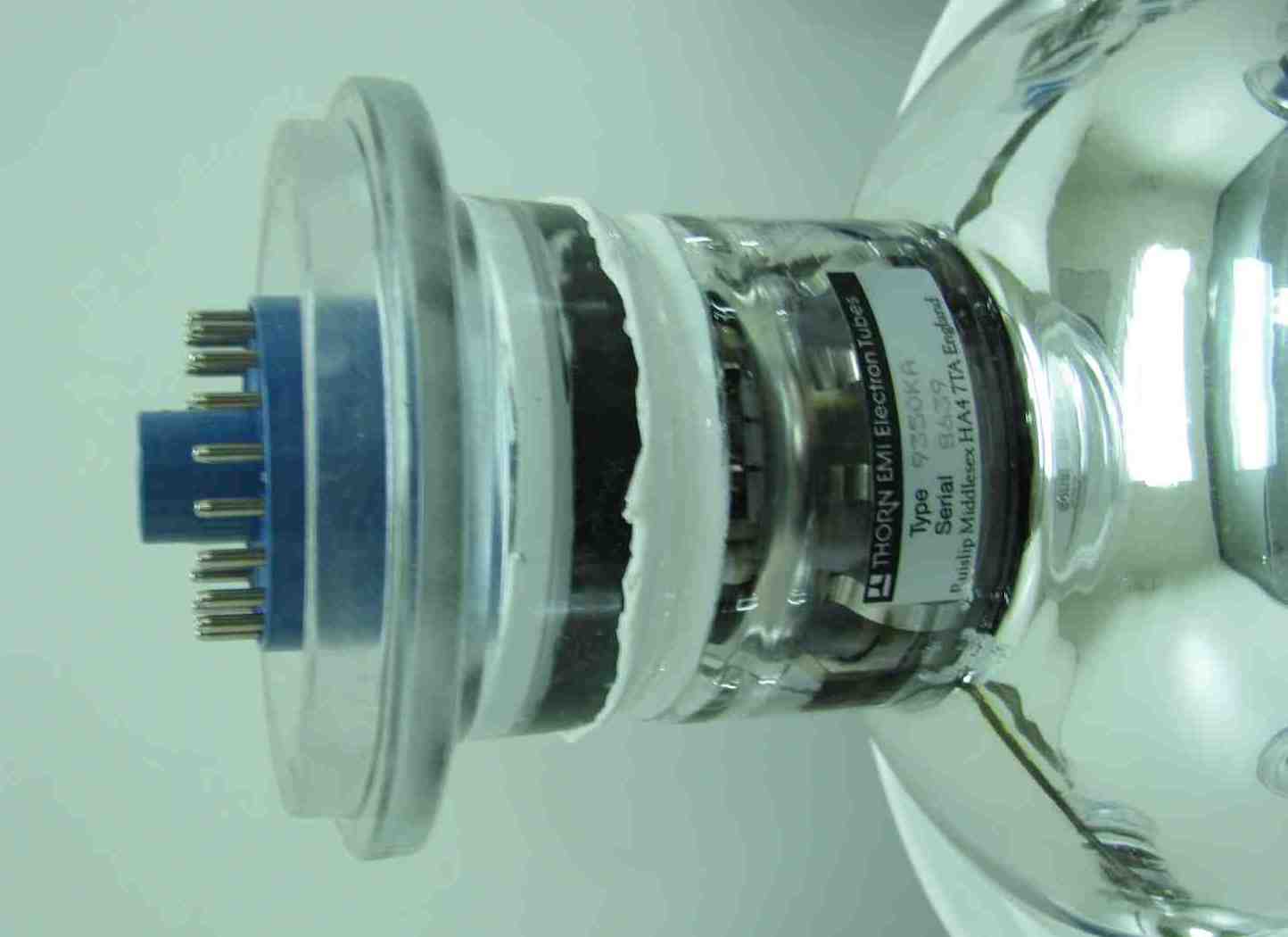} 
\includegraphics[width=60mm, height=45mm, angle=270]{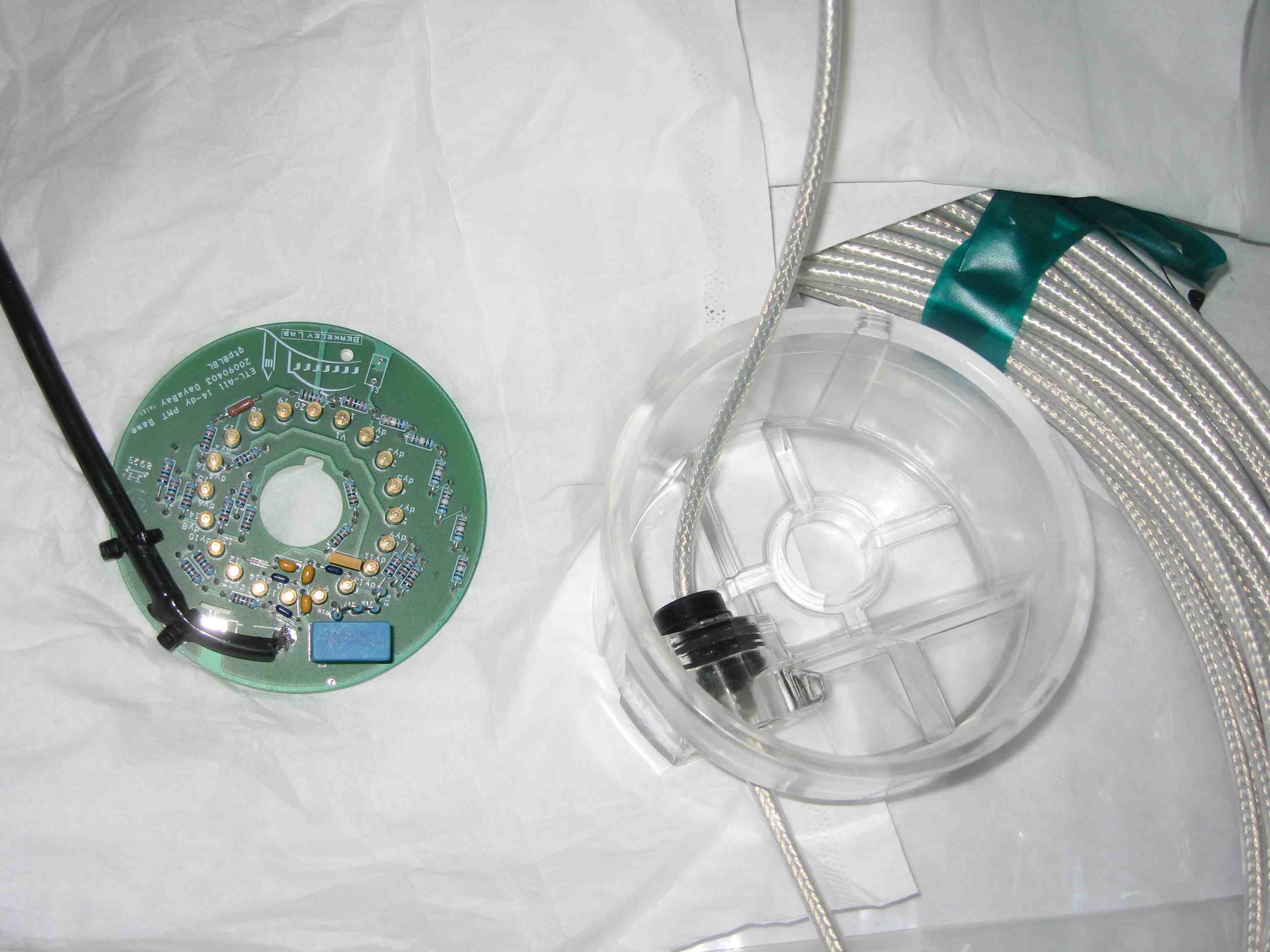} 
\caption{Left: pre-assembled MACRO PMT with an acrylic base. Mastic tape and marine sealant are applied to the cylindrical glass section of the PMT. Right: pre-assembled acrylic cap with cable and voltage-divider board.}
\label{fig:preAssembly}
\end{figure}

\subsection{Assembly}\label{sec:assembly}
Prior to the assembly, the acrylic shells were annealed in an oven to release any mechanical stress incurred in the molding process.
Assembly of the components was performed inside a laminar-flow fume hood to keep everything clean. The assembly consisted of several steps. The first step was to attach the acrylic base to the PMT. Mastic tape cut with a water jet was then applied to the cone-shaped region of the glass envelope at the proper height. A heat gun was used to soften the mastic tape beforehand to improve the sealing quality. After the mastic tape was applied, the acrylic base was pressed onto the mastic tape so that the glass, mastic tape and acrylic base adhered to each other tightly. The marine sealant was then used to fill the space between the acrylic 
base and the PMT socket. It was also applied around the outer perimeter, between the acrylic base and PMT glass. The cure time of the marine sealant was about 24 hours. A MACRO PMT with the acrylic base assembled is shown on the left side of Figure \ref{fig:preAssembly}. 

The next step was to assemble the PMT base with its HV cable and an acrylic cap. The HV cable was fed into the acrylic cap through the opening on the side wall. The etched end of the cable without an SHV connector was then soldered onto the voltage-divider board. The right picture of Figure \ref{fig:preAssembly} shows the result. 
The pre-assembled 
UW-PSL sealing plug on the cable was then pushed into the opening on the side wall. When the marine sealant on the acrylic base had cured and the polyethylene donut was placed the voltage-divider board was plugged onto the socket of the PMT. The acrylic base and acrylic cap were then joined with WeldOn-3 and WeldOn-16 acrylic cements, which have a cure time of less than 30 minutes. 
About an hour after the acrylic cements were applied the PMT was placed under a fume hood. 

\begin{figure}[htbp] 
\centering
\includegraphics[width=90mm, height=60mm]{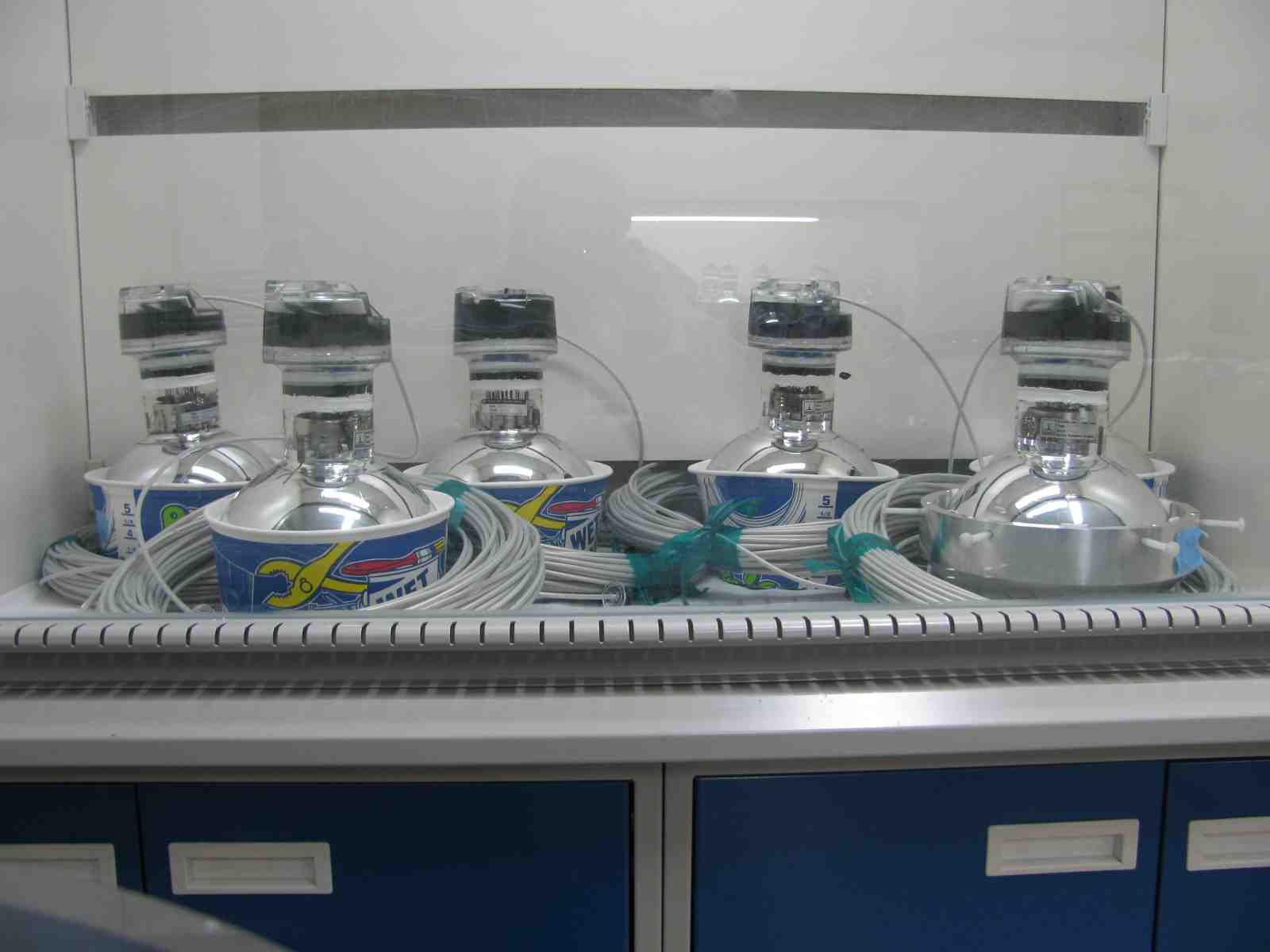} 
\caption{Potted PMT assemblies in the fume hood with a running exhaust fan.}
\label{fig:laminarHood}
\end{figure}

\subsection{Potting}\label{sec:lab}
The last step of the assembly was to seal the PMT base with epoxy. The two components of Epoxy 3M Scotch Cast SC2130 were mixed thoroughly and air bubbles were removed with a vacuum degassing chamber. The degassed epoxy was then poured into the acrylic shell through the top filling hole in the cap until the voltage-divider circuit board was immersed. The epoxy normally cured in 40 minutes. After four hours to a whole night, the filling hole was sealed with a small acrylic cap and acrylic cement. The opening with the UW-PSL sealing plug inserted was then sealed with a flat acrylic cap which also served as a strain relief for the cable. Once the acrylic cement had cured, mastic tape was applied to the region where the cable exited the acrylic cap. 
Figure \ref{fig:laminarHood} shows the completely assembled PMT units.\\

\subsection{Mechanical and Electrical Acceptance Tests}\label{sec:accep}

\subsubsection{Pressure Test}\label{sec:elec}
Each potted PMT was subject to a pressure test, mimicking the working conditions in the water pools at Daya Bay. The basic idea was to partially
submerge 
each potted assembly in de-ionized (DI) water inside a 47.3-liter pressure 
tank (maximum gauge pressure: $5.52 \times 10^5~Pag$). 
The tank was then pressurized to a pre-determined value for a period of
time. After the test was over, the assembly was inspected for water leakage. 

\begin{figure}[htbp] 
\centering
\includegraphics[width=60mm, height=50mm]{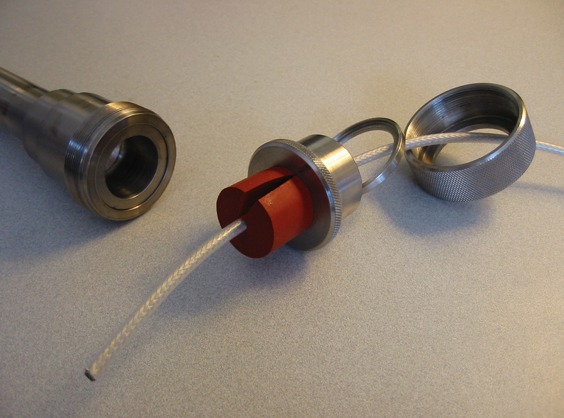} 
\includegraphics[width=60mm, height=50mm]{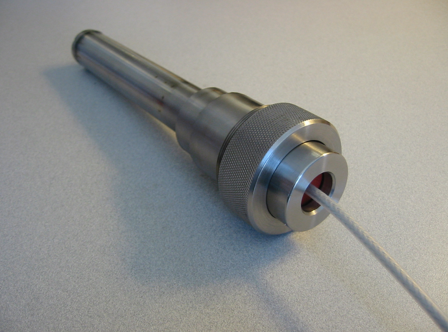} 
\caption{The cable end-sealing tube. Left: the components of the cable end-sealing tube. Right: the assembled cable end-sealing tube}
\label{fig:sealingTube}
\end{figure}

Prior to the pressure test we visually inspected all the seals of each potted PMT
assembly to ensure the sealants were completely dry and free of defects
such as small holes. \\

To make sure that no fluid could get into the assembly during the pressure test, the end of the coaxial cable with an SHV connector was encapsulated inside a 
custom-built cable end-sealing tube. The components of the sealing tube are shown in the left picture of Figure \ref{fig:sealingTube}. The right picture of Figure \ref{fig:sealingTube} shows the assembled sealing tube with the cable in place.\\

The potted PMT assembly was then installed in the pressure tank which was
partially filled with DI water. The potted portion of the PMT was submerged in 
water, with the glass bulb remaining in air. The cable end-sealing tube was placed on a plastic top with an indentation to prevent the tube from rolling. Once the tank was sealed, pressure was applied slowly and monitored with a pressure gauge on the lid. Usually, the pressure test lasted about 8 to 15 hours for each 
assembly. A subset of assemblies was tested for around 40 hours. After the
test was complete, the PMT assembly was visually inspected for any leaks or damage.\\ 

At the beginning of mass production a small number of assemblies 
failed the pressure test.
One mode of failure was infiltration of water into the cable end-sealing tube
due to damage in the fragile jacket of the cable. For such a case,  with
the cable immersed in a bath of soda water, the location of the damaged 
region was determined by looking for an electrically conducting path between 
the ground braid of the cable and the water. After the damaged area was patched 
up with mastic tape and covered with heat-shrink tubing and electrical tape, 
the pressure test was repeated to confirm that the cable was repaired. 
In addition, the screening test was 
revised to include an integrity check of each cable before assembly.  \\

Another mode of failure was fracturing of the glass envelope.  
As specified by the manufacturer, the maximum allowable pressure is
$ 101~kPag$. 
However, a few PMTs imploded at a pressure of $82.7~kPag$ during the test. 
A few other assemblies did not implode, but their glass dome cracked. 
For these cracked tubes, we measured the thickness of the glass. The top portion of the glass bulb was the thickest, reaching $3~mm$, but at about halfway up the bottom hemisphere from the dynode structure, the glass thickness
was about $1~mm$ or less. 
Accordingly, cracking usually occurred in the bottom hemisphere below the equator. This observation motivated us to weigh the bare PMTs before and after potting. The imploded or cracked PMTs 
were found to have a net weight of less than $540~g$. \\

\begin{figure}[htbp] 
\centering
\includegraphics[width=60mm, height=40mm]{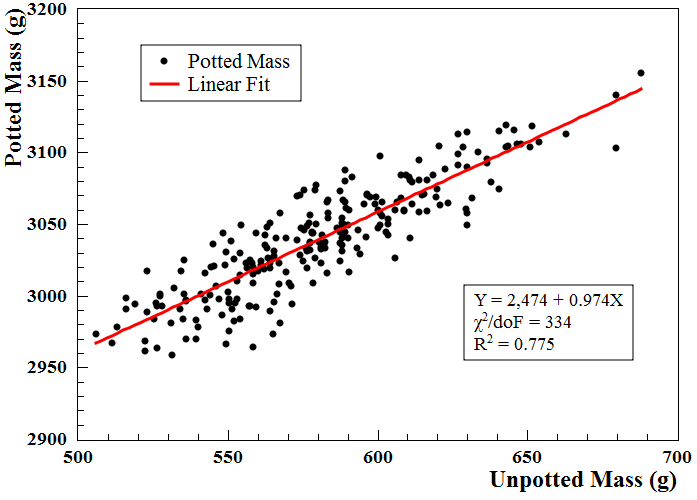} 
\includegraphics[width=60mm, height=40mm]{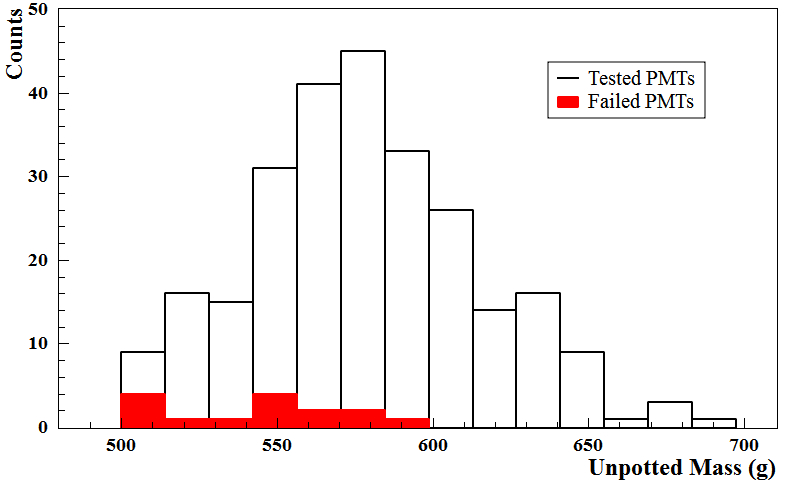} 
\caption{Left: Correlation between mass before and after potting, providing a means to determine a proper pressure for testing the potted assemblies. 
Right: Black (Red) histogram is the mass distribution of the potted tubes 
that passed (failed) the pressure test.}
\label{fig:tubeMass1}
\end{figure}

Since the PMT assemblies would be installed at different depths in the water pools, we decided to apply a lower pressure (P) for tubes with net mass (M) smaller than $540~g$. 
Specifically, if M $< 540~g$, P $= 41.4~kPag$; M $> 540~g$, P $= 82.7~kPag$. The potted MACRO assemblies that passed the test at $41.4~kPag$ were installed in the upper eighth of the water pools (static pressure $\leq 12~kPag$), and those that survived a pressure of $82.7~kPag$ were installed in the upper half of each pool ($< 41~kPag$). The lower halves of the water pools were populated exclusively with the new Hamamatsu PMTs.  Potted but unweighed MACRO tubes were temporarily put aside. After potting all the tubes, we found the correlation between the before- and after-potting masses (left plot of Figure \ref{fig:tubeMass1}). This correlation allowed us to estimate the net mass of the tubes 
that we did not weigh before potting, and then apply a proper pressure to test them. The right 
plot of Figure \ref{fig:tubeMass1} shows the distributions of the before-potting net mass of passed and failed assemblies. The tubes that broke during the test tend to have smaller mass. \\

~~During mass production, we potted eight tubes per day and performed pressure tests for two to three potted assemblies. By the end of March 2010, we had produced a total of 386 waterproofed assemblies, of which 52 were assembled before the potting and acceptance testing procedures were revised.  Of these, 50 were tested and 36 passed the pressure test, giving a success rate of 72$\%$. The remaining 334 tubes were assembled after the procedures of potting and pressure testing were refined. Among them, 190 were tested with 176 passed. The success rate improved to about 93$\%$. Six of these 176 waterproofed assemblies were sent to Rensselaer Polytechnic Institute for long-term pressure testing. In the end, we had 352 waterproofed PMT assemblies available for use, including 146 that had not been pressure-tested. These were later tested in the Surface Assembly Building at Daya Bay from August through December of 2010 with a success rate greater than 95$\%$.

\subsubsection{Electrical Test}\label{sec:elec}
A facility for testing all the PMT assemblies to be used in the Daya Bay 
experiment was set up at the Dongguan University of Technology (DGUT), which is about a two-hour drive from Daya Bay.
The 352 waterproofed MACRO PMT assemblies were sent to DGUT for electrical testing at the end of March 2010.  
\begin{figure}[htbp] 
\centering
\includegraphics[width=60mm, height=40mm]{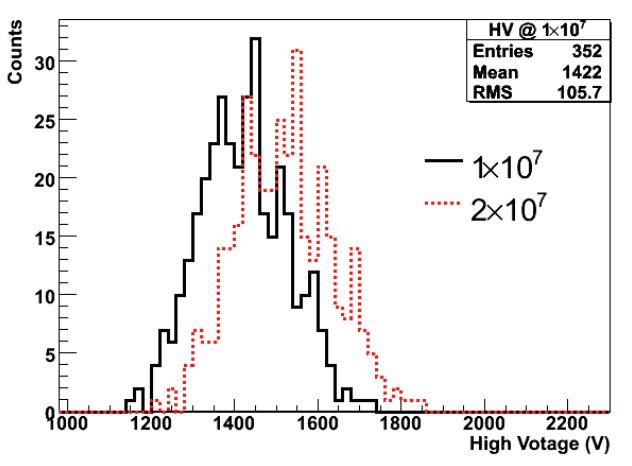} 
\includegraphics[width=60mm, height=40mm]{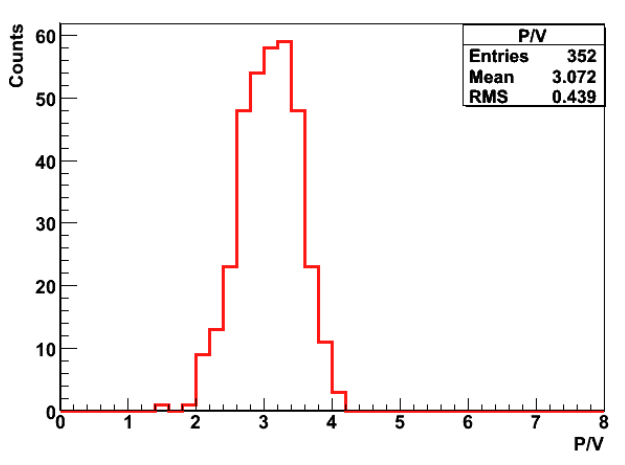} 
\includegraphics[width=60mm, height=40mm]{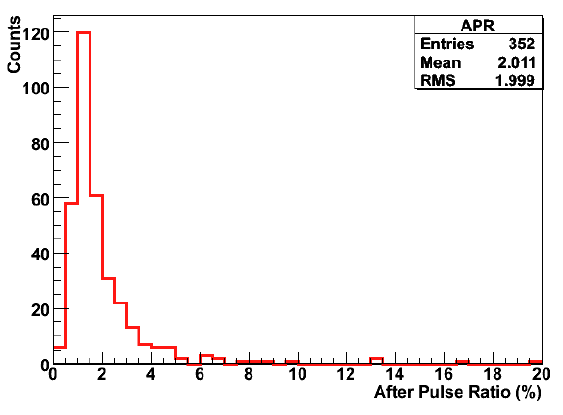} 
\includegraphics[width=60mm, height=40mm]{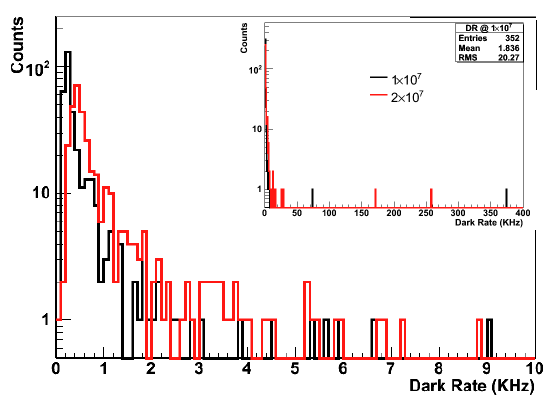} 
\caption{Top left: HV at the gain of 1$\times10^7$(black) and 2$\times10^7$(red); Top right: peak-to-valley ratio at a gain of 1$\times10^7$; Bottom left: after-pulsing ratio at the gain of 1$\times10^7$; Bottom right: dark rates at the gain of 1$\times10^7$(black) and 2$\times10^7$(red).}
\label{fig:testResults}
\end{figure}

The test included measuring the single photoelectron spectrum and its peak-to-valley ratio (P/V), gain as a function of the applied HV, dark rate, and after-pulsing ratio (APR). 
Here, APR is defined as the probability of observing a pulse excessing a threshold of 0.25 photoelectron within about 20 $\mu s$ after the main pulse.
Figure \ref{fig:testResults} shows the distributions of HV for operating the PMTs at gains of 1$\times10^7$(black) and 2$\times10^7$(red), P/V at the gain of 1$\times10^7$, APR at the gain of 1$\times10^7$ and the dark rate at the gains of 1$\times10^7$(black) and 2$\times10^7$(red), for the 352 MACRO PMTs. There were 17 assemblies that did not meet our requirements. Among them, one showed no output, two had P/V less than two, four had an APR larger than 10$\%$, and ten had dark rates higher than $10~kHz$ at gains of [0.3 - 3]$\times10^7$.  We decided that six of the ten noisy PMTs having 
dark rates just above $10~kHz$ and two others with P/V slightly lower than two could still be used. In total, there were 343 waterproofed MACRO PMTs available
to the experiment.

\subsection{Performance of Waterproofed PMT Assemblies at Daya Bay}\label{sec:waterpool} 
\begin{figure}[htbp] 
\centering
\includegraphics[width=60mm, height=40mm]{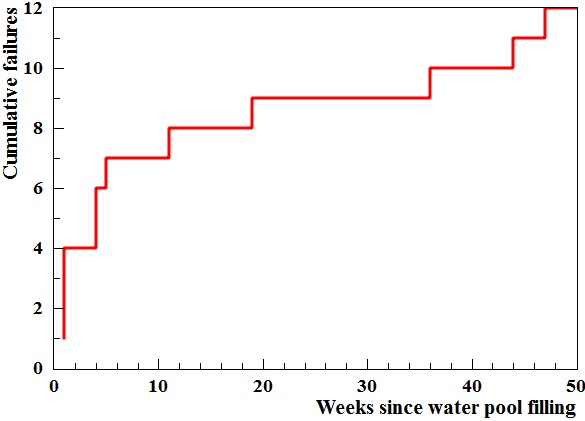} 
\includegraphics[width=60mm, height=40mm]{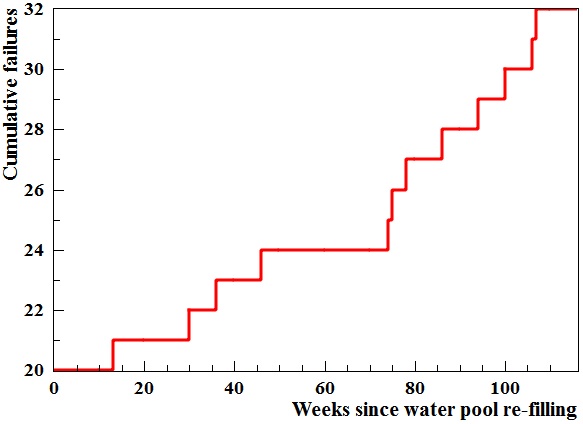} 
\caption{Cumulative number of failed installed PMT assemblies in all experimental halls during the first year (left) and
subsequent two years (right) of data taking. 
Note that zero is suppressed in the y-axis for the graph shown on the right.
\label{fig:tubeFail}}
\end{figure}

By the end of 2011, all waterproof PMT assemblies were installed and commissioned in the three experimental sites. Among them, 104 of the 288 in both EH1 and EH2, and 133 of the 384 in EH3 were waterproofed MACRO PMTs. Taking EH1 as an example, the singles rates of the PMTs were 2-8 $kHz$ in the inner partition of the water pool, and 10-16 $kHz$ in the outer one. There was no noticeable difference in the singles rates between the new Hamamatsu and refurbished MACRO waterproof PMTs~\cite{bib:muonSys}. 
Over the course of nearly one year of data-taking, a number of the waterproofed assemblies failed. The cumulative number of failures over this time is shown in the left plot of Figure \ref{fig:tubeFail}. 
The mass distribution of the PMTs that failed in the water pools is very similar to that of the PMTs that failed the pressure test during fabrication, the latter of which is shown in Figure \ref{fig:tubeMass1}.  \\

From the end of July through September 2012, steady data-taking was paused to perform special AD calibration measurements and install the last two ADs, one in EH2 and one in EH3.  During this time, the failed PMTs in the water pools were investigated and all but one were replaced with spare Hamamatsu PMTs.  In total, 6, 5, and 2 replacements were made in EH1, EH2, and EH3 respectively. The failed PMTs were tested for cable leaks using a custom-made cable/air pressurization tube: no cable leak was found.  All but two of the failed PMTs were found to have cracks in the glass extending from the equator of the dome near the support ring upward away from the base.  One of the two remaining failed PMTs imploded after data-taking was paused and it was removed from EH1, while the other was removed from EH2 because it had a high dark rate that might have been caused by a leak in the base.  Upon refilling the pools, 2 and 6 new failures occurred in EH2 and EH3, respectively.  It is hypothesized that the changes in pressure across the PMT glass during draining and filling the pools triggered the failures.  It was decided not 
to re-drain the pools and replace these new failures due to the risk of causing more failures. After all, the very small number of such failures has minimal impact on 
the performance of the water Cherenkov counters.  In the subsequent two years of steady data-taking, 20 additional MACRO PMT assemblies failed in the three pools (see the right plot of Figure \ref{fig:tubeFail}). 
Since October 2012 (corresponding to 70 weeks in Figure \ref{fig:tubeFail}), the failure rate appears 
to have stablized to $\leq$0.5 PMT/month. 

\section{Summary}	
We have successfully fabricated 386 waterproofed PMT assemblies by recycling some of the MACRO PMTs. The motivation of this effort was to avoid the expense of purchasing new waterproof PMT assemblies. However, the complexity and cost of this effort were, in the end, higher than expected. We have installed 341 of these waterproofed PMTs in the muon system of the Daya Bay experiment, corresponding to slightly more than one third of all PMTs in the water pools. During the first twelve months of operation, 15 of these 341 PMTs malfunctioned in the water pools but only one is thought to have failed due to the imperfect sealing of the PMT base.  The electrical performance of the waterproofed MACRO PMTs has been very similar to that of the new Hamamatsu waterproof R5912 PMTs.

\vspace{0.5 cm}
{\bf Acknowledgement} \\
\vspace{0.15cm}

We would like to thank Istituto Nazionale di Fisica Nucleare (INFN) for offering the MACRO PMTs. 
We are grateful for the support provided 
by the Director, Office of Science, Office of High Energy Physics, of the U.S. 
Department of Energy under Contract No. DE-AC02-05CH11231. 

\end{document}